\documentclass[doublecol]{epl2} 
\usepackage[british]{babel}
\usepackage[shortcuts]{extdash}
\usepackage{amssymb}
\usepackage{amsmath}
\usepackage{bbm}

\renewcommand{\revision}[1]{#1}

\renewcommand{\Re}{{\mathop{\mbox{Re}}}}
\renewcommand{\Im}{{\mathop{\mbox{Im}}}}
\newcommand{\tr}{{\mathop{\mbox{tr}}}}
\newcommand{\proP}{{\mathcal P}}
\newcommand{\proQ}{{\mathcal Q}}
\newcommand{\id}{{\mathbbm{1}}}
\newcommand{\Liou}{{\mathcal L}}
\newcommand{\tmem}{\tau_{\rm mem}}
\title{Exact propagation of open quantum systems in a
system-reservoir context}

\author{J\"urgen. T. Stockburger\inst{1}}
\shortauthor{J. T. Stockburger}

\institute{                    
  \inst{1} Institute for Complex Quantum Systems, Ulm University
 - Albert-Einstein-Allee 11, 89081 Ulm}

\abstract{A stochastic representation of the dynamics of open quantum
  systems, suitable for non\-/perturbative system-reservoir
  interaction, non-Markovian effects and arbitrarily driven systems is
  presented. It includes the case of driving on timescales comparable
  to or shorter than the reservoir correlation time, a notoriously
  difficult but relevant case in the context of quantum information
  processing and quantum thermodynamics. A previous stochastic
  approach is re-formulated for the case of finite reservoir
  correlation and response times, resulting in a numerical simulation
  strategy exceeding previous ones by orders of magnitude in
  efficiency.  Although the approach is based on a memory formalism,
  the dynamical equations propagated in the simulations are
  time-local. This leaves a wide range of choices in selecting the
  system to be studied and the numerical method used for
  propagation. For a series of tests, the dynamics of the spin-boson
  system is computed in various settings including strong external
  driving and Landau-Zener transitions.}

\begin{document}

\maketitle

\section{Introduction}

The reduced density matrix of an open quantum system is the
fundamental mathematical object characterizing the system's state and
dynamics. It is defined through a partial trace operation on the
density matrix of a larger system, often referred to as ``tracing out
the environment''. Finding a suitable description of dynamics for a
state which is thus, by definition, an \emph{incomplete} description
of physical reality becomes a non-trivial task, for which a number of
techniques, most of them approximate, have been developed.

Quantum master equations have been used successfully
where the influence of an environment on the dynamics can be
characterized as perturbative and Markovian. Master equations of
Lindblad form \cite{alick87,breue02} are often preferred since they
generate completely positive maps.

Lindblad terms not only reflect properties of the environment and the
system-environment interaction, but also the dynamics and level
structure of the \emph{system}, to the degree it can be resolved
within time intervals of the order of the correlation time of the
environment. In the typical case of an equilibrated environment with
thermal energy $k_{\rm B}T$ lower than the system's level spacing, the
dependence of Lindblad operators on the specific properties of the
system Hamiltonian is crucial: The thermal timescale $\hbar\beta$ is
long enough to differentiate between the energy levels of the system,
making the approach difficult to apply to complex or driven
systems. This applies in particular when driving is not adiabatic on
the timescale of reservoir fluctuations \cite{alick06}. When driving
parameters are chosen such that the level structure of the system
changes appreciably over time intervals of width $\hbar\beta$ (thermal
timescale), novel effects appear \cite{schmi11} which are not within
the scope of standard master equations.

Sometimes ad-hoc combinations of Hamiltonian and Lindbladian terms are
used to \emph{define} a dissipative system. While this approach is
guaranteed to result in a completely positive channel, there are
important cases where it is impossible to reconcile with the full
coupled dynamics of the system coupled to a thermal reservoir
\cite{levy14,stock16}.

\revision{%
Non-perturbative approaches to open-system dynamics include formally
exact dissipative path integrals \cite{feynm63,weiss12} as
well as renormalization group methods
\cite{kehre97,keil01,ander07}, which provide a computational approach
suited to shed light on quantum phase transitions.

Dissipative path integrals contain a non-local action term
\cite{feynm63}, which makes it difficult to find a fully equivalent
equation of motion for the reduced density matrix. Several strategies
have been employed, among them the} time-discrete propagation of a
multidimensional tensor state \cite{makar94}, equations of motion
for a hierarchy of auxiliary density matrices \cite{tanim89}, and
mapping environmental fluctuations and response on stochastic
processes. Several stochastic approaches are known, among them the
non-Markovian extension of quantum state diffusion \cite{diosi97},
stochastic decoupling of system and reservoir \cite{shao04,lacro05}
and the stochastic unravelling of influence functionals
\cite{stock02,stock04}. The latter approach is applicable in the
generic case of linear dissipation with arbitrary spectral
characteristics. In the following, a variant of this approach will be
developed which shares features with existing finite-memory approaches
to reduced dynamics \cite{golos99,cerri14} but leads to equations of
motion which are time-local and allow arbitrary time-dependent
Hamiltonians.

The finite-memory stochastic propagation (FMSP) variant of the
stochastic approach presented here shows improvements in efficiency
reducing the required computational resources by orders of
magnitude. In particular, the revised method is now suitable to fully
explore equilibration processes, non-equilibrium steady states and the
long-time limit of cyclic processes.

\section{Stochastic Liouville--von Neumann equation}

The path integral representation \cite{feynm63} of an open quantum
system describes the effects of coupling to an environment in terms of
an influence functional, a functional of a pair of forward and
backward paths which depends on properties of the system-reservoir
coupling, the reservoir dynamics and the initial state of the
reservoir. It does not depend on the system's intrinsic (or forced)
dynamics in any way, however, it is not a time-local functional. It
cannot be transformed into a deterministic equation of motion for the
reduced density matrix unless quite restrictive approximations are
made or a significant number of auxiliary dynamical states are
included \cite{tanim89,makar94}.

In the case of linear dissipation, i.e., Gaussian free fluctuations of
the reservoir, the influence functional itself is also a Gaussian
functional. Here we consider a system-reservoir interaction
$H_{\rm I} = - q\cdot B$, where $q$ is a system coordinate and $B$ is
typically a force depending on many reservoir coordinates, with an
obvious generalization to arbitrary sums of separable terms.

The resulting Feynman-Vernon influence functional \cite{feynm63} is
completely characterized by the complex-valued correlation function
\begin{equation}
L(t-t') = \frac{1}{Z} \tr\left( \exp(-\beta H_{\rm R}) B(t)B(t')\right)
\end{equation}
of the free reservoir
fluctuations. The function $L(t-t')$ describes both fluctuations (real
part) and dynamical response (back action, imaginary part).

The mathematical structure of a Feynman-Vernon influence functional is
closely related to generating functionals of stochastic processes
governed by \emph{classical} probability. This allows a stochastic
re-formulation of open-system dynamics in terms of the stochastic
Liouville--von Neumann equation \cite{stock02,stock04}
\begin{equation}\label{eq:SLN}
i\hbar \frac{d}{dt}{\rho} = \Liou \rho
 = [H_{\rm S},\rho] - \xi[q, \rho] - \frac{\hbar\nu}{2}
[q, \rho]_+
\end{equation}
With its time evolution governed by two correlated stochastic
process $\xi(t)$ and $\nu(t)$, the dynamical state $\rho$ itself
becomes a stochastic variable; the \emph{physical} reduced density matrix
\begin{equation}\label{eq:SLNav}
\bar\rho(t) = \langle\rho(t)\rangle
\end{equation}
is obtained as the expectation value of samples $\rho(t)$. Note that
angle brackets $\langle\cdot\rangle$ refer to stochastic averages
throughout this paper, no trace operation is implied.

For eq. (\ref{eq:SLN}) to match the original system-reservoir model,
the following conditions are sufficient \cite{stock04}:
\begin{align}
\langle\xi(t)\xi(t')\rangle &= \Re L(t-t')\nonumber\\
\langle\xi(t)\nu(t')\rangle &= (2i/\hbar)
\Theta(t-t')\Im L(t-t') + i\mu \delta(t-t')\nonumber\\
 &= -i \chi(t-t') + i \delta(t-t')
  \int_0^\infty d\tau \chi(\tau) \delta(t-t') \nonumber\\
\langle\nu(t)\nu(t')\rangle &= 0 \label{eq:simplenoise}
\end{align}
These conditions can be fulfilled (and corresponding noise samples
generated using the fast Fourier transform method) \emph{provided}
that $\xi(t)$ and $\nu(t)$ are allowed to take complex values. The
mathematical result (\ref{eq:SLN})--(\ref{eq:simplenoise}) is
physically counter-intuitive on the level of individual samples: There
is no response or damping term in eq. (\ref{eq:SLN}); the dynamic
response function $\chi(t-t')$ emerges only after averaging
over noise realizations.

The preceding equations are universally valid for arbitrary
dissipation strength and spectral characteristics of the reservoir
fluctuations. Unlike Lindblad terms, the stochastic terms \revision{in}
eq. (\ref{eq:SLN}) do not depend on $H_{\rm S}$ in any way; it is
perfectly legitimate to modify the Hamiltonian $H_{\rm S}$ in
(\ref{eq:SLN}) while keeping the stochastic terms unchanged. In
particular, the stochastic Liouville--von Neumann equation can
accommodate external driving with arbitrary time dependence, which
would break the assumptions underlying the standard derivation of
commonly used master equations.

Efficient simulation methods based on slight modifications of
eq. (\ref{eq:SLN}) have been developed for semiclassical dynamics
\cite{koch08} and for continuous degrees of freedom with ohmic
friction \cite{stock99}. In the general case, however, the numerical
cost of averaging eq. (\ref{eq:SLN}) over explicitly drawn samples of
$\xi$ and $\nu$ can be prohibitive. Eq. (\ref{eq:SLN}) contains $\xi$
and $\nu$ as \emph{multiplicative} noise, leading to asymptotic
long-time behaviour similar to geometric Brownian motion, the
prototypical example of multiplicative noise. Empirical data indicate
that the second moment of $||\rho||$ (Frobenius norm) exists for
arbitrary $t$, but grows exponentially in the asymptotic regime of
large $t$. While the method has been found valuable as a computational
tool for transient phenomena, even in a system as complicated as the
FMO bacteriochlorophyll complex \cite{imai15}, its numerical cost in
the long-time limit grows exponentially. The sequel illustrates a
refined approach which overcomes this problem.

\section{Finite-memory stochastic propagation}

A frequent situation in the dynamics of an open quantum system
interacting with a reservoir is the following: The reservoir
correlations have a finite correlation time, yet this time is not
short enough to allow a Markovian approximation. Hence, whenever
$L(\tau)$ is effectively a function with finite support,
modified simulation strategies based on eq. (\ref{eq:SLN}) can be
found with much more benign requirements of computational resources.

Forming the expectation value $\bar\rho = \langle\rho\rangle =
\proP\rho$ can be viewed as the definition of a projection operation
$\proP$, with complement $\proQ= \id - \proP$. Eqs. (\ref{eq:SLN}) and
(\ref{eq:SLNav}) are thus related to a Nakajima-Zwanzig
equation
\begin{equation}
\frac{d}{dt} \bar\rho = \proP\Liou  \bar\rho
 + \proP\Liou  \int\limits_{t^*}^t dt'
\exp_>\left(\proQ\Liou  (t-t')\right)\proQ\Liou  \bar\rho(t')\;,
\label{eq:zwanzig}
\end{equation}
which is a formal equation of motion for the relevant part $\bar\rho = \proP
\rho$, with $\exp_>$ denoting a time-ordered exponential.

With $t^*=0$ and the initial condition $\proQ\rho(0) = 0$, this
equation is exactly equivalent to eqs. (\ref{eq:SLN}) and
(\ref{eq:SLNav}). Now it is important to note that the lower integration
boundary can be raised to $t-\tmem$ without incurring noticeable
errors in the case of finite memory time $\tau_{\rm mem}$ of the
environmental effects.

With $\proP$ in the original meaning of ``tracing out the
environment'', time-discrete versions of eq. (\ref{eq:zwanzig}) with
finite memory time have recently been discussed
\cite{golos99,cerri14}, with a focus on explicitly determining the
discrete analogue of the memory operator $\proP\Liou \exp_>\left(\proQ\Liou
(t-t')\right)\proQ\Liou$.

The finite-memory case $t^* \lessapprox t-\tmem$ also benefits the
stochastic approach, since the ``noisy'' propagator $\exp_>(\proQ\Liou
(t-t'))$ is then applied only to time intervals of length up to
$\tmem$ rather than $t$ (in the infinite-memory case, $t^*=0$).

An efficient simulation algorithm with greatly improved sampling
statistics results from the following approach: Instead of computing
the integral in (\ref{eq:zwanzig}) for each time step of a numerical
solution, it is advantageous to transform eq. (\ref{eq:zwanzig}) back
into a system of differential equations for the relevant part
$\bar\rho = \proP\rho$ and the irrelevant part $\breve\rho =
\proQ\rho$. However, multiple instances of $\breve\rho$ are needed
since different lower bounds of the integral amount to different
initial conditions for $\breve\rho$.

It is therefore favourable to choose the lower integration boundary as
a piecewise constant staircase function $t^*(t)$, bounded by $t-\tmem$
from above and by $t-\tmem-\tau^*$ from below. The ``staircase
timescale'' $\tau^*$ is an intermediate time scale shorter than
$\tmem$, to be discussed in further detail below.

One thus arrives at the equivalent set of coupled equations
\begin{align}
\dot{\bar\rho} &= \proP \Liou\bar\rho + \proP\Liou \breve\rho_{m(t)}
\label{eq:barrho}\\ 
\dot{\breve\rho}_n &= \proQ \Liou\bar\rho  + \proQ\Liou \breve\rho_n
\label{eq:breverho}
\end{align}
with the initial conditions $\bar\rho(0) = \rho_0$ at $t=0$ and
$\breve\rho_n(n\tau^*) = 0$ at equidistant times $n\tau^*$. Equation
(\ref{eq:barrho}) is a linear deterministic equation, with an
inhomgeneous term $\proP\Liou \breve\rho_{m(t)}$ containing the covariance
$\langle \xi\breve\rho_{m(t)}\rangle$, while eq. (\ref{eq:breverho})
is a linear stochastic equation with inhomogeneity $\proQ
\Liou\bar\rho$ containing products of $\bar\rho$ with either noise
variable.

The index $m(t)$ is given by $m(t) =
\max(0,\lfloor(t-\tmem)/\tau^*\rfloor)$, where the brackets $\lfloor
\cdot \rfloor$ denote the floor function. This ensures that the time
interval between the initialization of $\breve\rho_{m(t)}$ and its
first appearance in eq. (\ref{eq:barrho}) is longer than
$\min(t,\tmem)$.

\section{General notes on implementation}

Formally, eq. (\ref{eq:breverho}) suggests an infinite set of
equations. However, at any time $t$ those $\rho_n$ with
$n>t/\tau^*$ have not even been assigned their initial values, and all
$\rho_n$ with $n<(t-\tmem)/\tau^*$ can safely be discarded since
$m(t)\neq n$ for all future times. The number of ``active'' instances
$\breve\rho_n$ being propagated at any given time $t$ is finite, it is
bounded by $\tmem/\tau^*+1$.

The major benefit of this procedure lies in the repeated averaging of
\emph{partial} results in the course of propagating
eq. (\ref{eq:barrho}) as well as in the re-setting of the initial
state of $\breve\rho$: As a consequence of this, the \emph{growth
  rates} of the variances of $||\bar\rho||$ and $||\breve\rho||$
become arbitrarily small when the number of samples is
increased, i.e., the number of samples needed for given $t$ no longer
grows exponentially.

The timescale $\tau^*$ is chosen to roughly satisfy two
criteria: The number of active instances $\approx \tmem/\tau^*$ should
not be too large, and the maximum effective memory time $\tmem +
\tau^*$ should not be much larger than $\tmem$. The memory time
$\tmem$ itself is typically chosen as a multiple of the reservoir
correlation time to account for higher-order processes.

Any numerical propagation scheme capable of dealing with equations of
type (\ref{eq:barrho}) and (\ref{eq:breverho}) is compatible with this
approach; in particular, its time step can be chosen independently of
$\tau^*$. In the following examples, a simple split-operator technique
is used.

\section{Numerical examples}

\begin{figure}
\begin{center}
\includegraphics[width=\columnwidth]{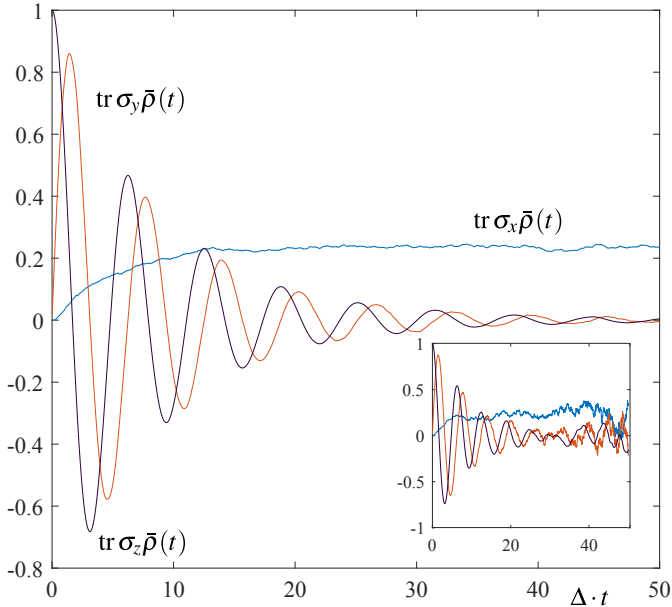}
\caption{Performance test: Spin-Boson dynamics, ohmic dissipation with
  dissipation constant \cite{weiss12} $K = 0.2$, thermal time (scaled
  inverse temperature) $\hbar\beta = 1/2 \Delta^{-1}$ and $\omega_c =
  10\Delta$. Inset: direct simulation of eq. (\ref{eq:SLN}), same
  parameters. See text for simulation details.\label{fig:SB}}
\end{center}
\end{figure}

\begin{figure}
\begin{center}
\includegraphics[width=\columnwidth]{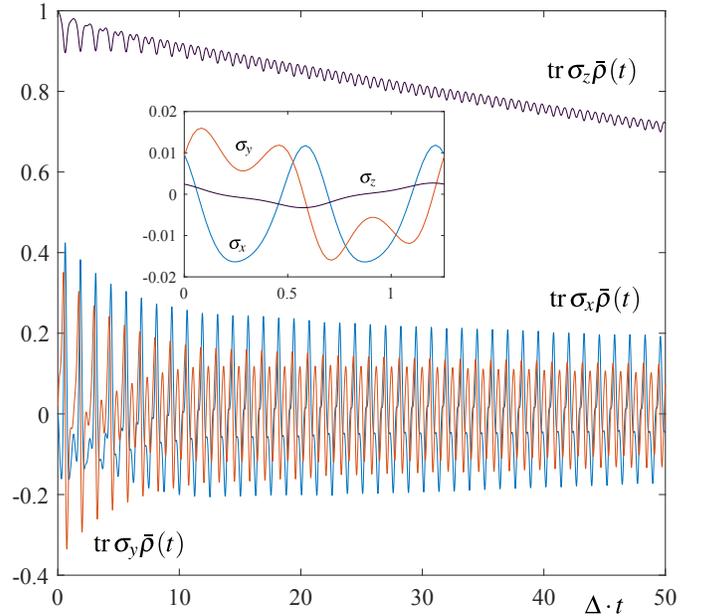}
\caption{Test case: Spin-Boson dynamics with high-frequency driving,
  ohmic dissipation with parameters $K = 0.02$, $\hbar\beta = 1/2
  \Delta^{-1}$ and $\omega_c = 10 \Delta$ and driving with parameters
  $\epsilon_1 = 12 \Delta$, $\omega_0 = 5\Delta$. \revision{Inset: asymptotic
  periodic dynamics (scaled time $\Delta\cdot t$ taken modulo driving
  period).}}
\label{fig:CDT}
\end{center}
\end{figure}

\begin{figure*}
\begin{center}
\includegraphics[width=0.95\textwidth]{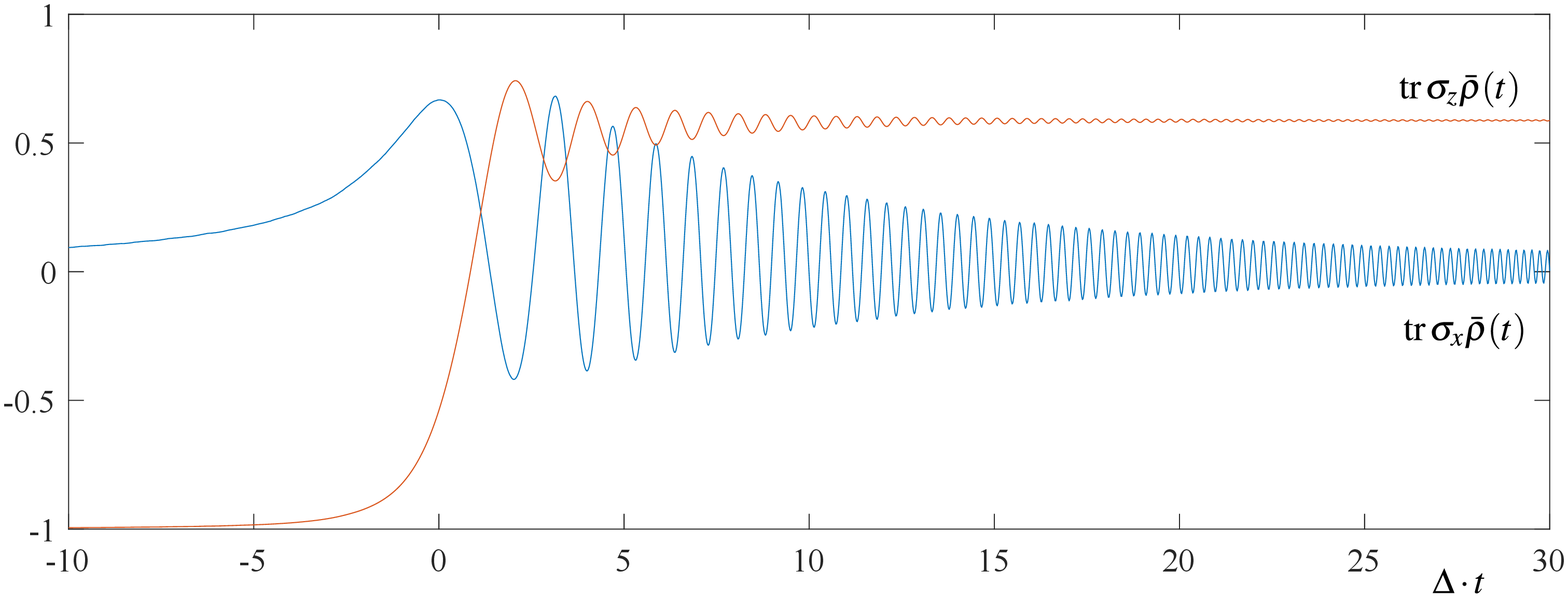}
\caption{Dissipative Landau-Zener dynamics: Dephasing completes the
  transition. Parameters are $v = \Delta^2$, $K = 0.05$,
  \revision{$\hbar\beta=2\Delta^{-1}$} and $\omega_c = 10\Delta$. Note
  that the decay of $\tr \sigma_x \bar\rho$ is
  non-exponential.}\label{fig:LZ}
\end{center}
\end{figure*}

The spin-boson model \cite{legge87,weiss12} is a well-studied, yet
non-trivial model which will be used to test the performance and
versatility of FMSP. In the simplest case,
the system part of the spin-boson Hamiltonian is a pseudo-spin subject
to a field in $x$ direction. The Hamiltonian for system, coupling and
\revision{bosonic reservoir can be written as
\begin{equation}\label{eq:SB}
H = -\frac{\hbar\Delta}{2} \sigma_x
- \sigma_z B
+ \sum_j \hbar\omega_j
a^\dagger_j a_j
\;,
\end{equation}
where $\sigma_z$ takes the role of $q$ in the preceding formalism, and
where $B = \sum_j \lambda_j (a_j + a_j^\dagger)/2$.} In the
case of ohmic friction, the reservoir is characterized by a
dimensionless dissipation constant K, the reservoir temperature, and a
spectral density\revision{%
\begin{equation}
G(\omega) =\pi\sum_j \lambda_j^2 \delta(\omega-\omega_j) = 2 \pi K
\omega f(\omega/\omega_c)
\end{equation}}
which rises linearly over a wide range of frequencies $\omega$ and
vanishes in the ultraviolet limit $\omega \gg \omega_c$. The choice \revision{of
an algebraic function} $f(x) = (1+x^2)^{-2}$ ensures rapid decay of
$\Im L(\tau)$ in the time domain.

Fig. \ref{fig:SB} shows simulation data obtained using FMSP with 500
samples. The system is initially prepared in an eigenstate of
$\sigma_z$, uncorrelated with the environment. $\tr \sigma_z \bar\rho$
and $\tr \sigma_y \bar\rho$ show the expected damped coherent
oscillations, while $\tr \sigma_x \bar\rho$ relaxes towards
equilibrium. The inset shows the result of a direct simulation of
eq. (\ref{eq:SLN}) with an equal number of samples.

What is noteworthy about this first test case may not be the physical
result per se, but the comparison of the error characteristics of the
two approaches. The statistical errors for the observables $\sigma_j$
and the Frobenius norm of $\breve\rho_n$ level out at plateaus at or
below an absolute value of $\approx 0.005$. The direct simulation of
eq. (\ref{eq:SLN}) shows a rapidly deteriorating signal-to-noise ratio
at long times (inset); its variance at $t=50 \Delta^{-1}$ is about two
orders of magnitude larger then for the finite-memory approach. Beyond
that time it grows exponentially with a rate roughly equal to the
absolute value of the dissipative decay rate.

We now turn to strongly driven spin-boson dynamics, fig. \ref{fig:CDT},
characterized through the Hamiltonian
\revision{%
\begin{equation}\label{eq:driven}
H = -\frac{\hbar\Delta}{2} \sigma_x
 + \frac{\hbar\epsilon(t)}{2} \sigma_z
- \sigma_z B
+ \sum_j \hbar\omega_j
a^\dagger_j a_j
\end{equation}}
with $\epsilon(t) = \epsilon_1 \cos(\omega_0 t)$, $\epsilon_1 =
12\Delta$ and $\omega_0 = 5\Delta$. Apart from rapid oscillations, the
dynamics shows a dramatic slow-down of relaxation and dephasing, which
is expected, since the driving parameters are near the regime of
coherent destruction of tunnelling
\cite{gross91a,stock99a}. \revision{%
The inset
of fig. \ref{fig:CDT} shows the asymptotic periodic dynamics,
determined from a separate run with symmetric initial condition,
extending up to $\Delta\cdot t = 2\cdot 10^2$.}

The stochastic construction (\ref{eq:simplenoise}) being completely
independent of the system Hamiltonian $H_{\rm S}$, all simulation
parameters relating to the dissipation mechanism (coupling strength,
noise spectra) are exactly the same as in fig. \ref{fig:SB}, except
that the number of samples has been increased by a factor of four to
allow resolution of the finer high-frequency features.

As a model case of non-periodic driving, a Landau-Zener transition is
considered next, i.e., driving of the form of a linear sweep
$\epsilon(t) = v t$. The dissipative Landau-Zener problem has a known
solution for zero temperature \cite{wubs06} and has been studied
numerically, using a related stochastic method adapted particularly to
ohmic dissipation \cite{orth10}. With a sweep speed $v=\Delta^2$,
fig. \ref{fig:LZ} shows both the change in population of $\sigma_z$
eigenstates and the gradual loss of coherence (decay of
$\tr \sigma_x \bar\rho$) which finalizes the transition. The
simulation is based on 30,000 noise samples and covers the interval
$\Delta\cdot t\in [-30,30]$. The decay of coherence is somewhat slower
than exponential; this is to be expected since the decay mechanism
itself is affected by driving. A simulation over the entire interval
covered here would have been infeasible using direct sampling of
eq. (\ref{eq:SLN}).

\begin{figure}[!bt]
\begin{center}
\includegraphics[width=\columnwidth]{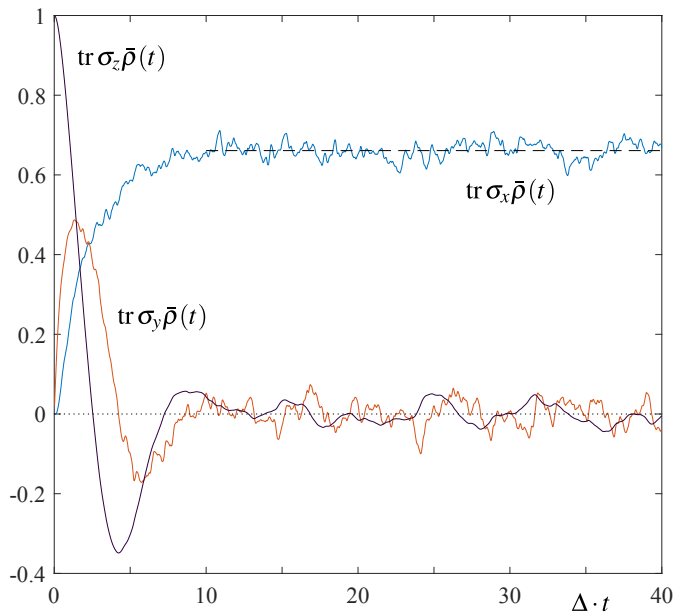}
\caption{Spin-Boson dynamics with \revision{moderate} ohmic dissipation,
  $K=0.2$, and a thermal time $\hbar\beta = 5 \Delta^{-1}$. Both time
  evolution and stationary state differ significantly from predictions
  of the Born-Markov approximation. The dashed line indicates the
  equilibrium value of $\tr \sigma_x \bar\rho \approx 0.66$
  estimated from a time average in the interval $\Delta\cdot t \in
  [10,40]$.}\label{fig:K2}
\end{center}
\end{figure}

The thermal timescale of a reservoir is often orders of magnitude
larger than its timescale of dynamic response; in the case of an ohmic
reservoir these would be $\hbar\beta$ and $1/\omega_c$. It is possible
\cite{stock04} to split the noise $\xi(t)$ into two independent parts,
$\xi(t) = \xi^{(s)} + \xi^{(l)}$, where $\xi^{(s)}$ is complex-valued
with a short correlation time $1/\omega_c$, and $\xi^{(l)}$ is
real-valued with correlation time $\hbar\beta$. The probability space
of the noise functions is thus a product of two independent spaces:
one with long-range, real-valued noise $\xi^{(l)}$ and one with
short-range, complex-valued noise $(\xi^{(s)},\nu)$. Applying
finite-memory propagation to the short-range part, while directly
sampling the long-range part, becomes attractive when stronger
dissipation and lower temperatures are to be considered. Simulation
results obtained with this approach are shown in fig. \ref{fig:K2}. It
is evident by visual inspection that the numerical error remains
roughly constant \emph{after} the system has equilibrated. This
variant of FMSP can therefore be used to gather thermodynamic
information from dynamical simulations. Stationary states may be
interpreted as equilibrium states in the case of non-perturbative
system-reservoir interactions, or, in the case of more than one
reservoir, non-equilibrium steady states. Taking time averages over
the stationary part of the dynamics, indicated by the dashed line,
further reduced statistical errors.

The expectation values $\tr \sigma_j \bar\rho$ provide a complete
parameterization of the reduced density matrix of the two-state
system. Due to symmetry, only $\tr \sigma_x \bar\rho$ is non-zero in
the stationary state. Its numerical value leads to quantitative
thermodynamic data, e.g., the entropy $S \approx 0.455 k_{\rm B}$, an
order of magnitude larger than the entropy $S \approx 0.0402 k_{\rm
  B}$ in the weak-coupling limit, where the occupation of the upper
level is less than 2 percent. This strong discrepancy can easily be
understood in a two-state model: For stronger coupling, there is a
competition between environment-induced superselection \cite{zurek03}
and local equilibration. Decreasing coherence between eigenstates of
$\sigma_z$ through decoherence automatically leads to a decrease of
the population difference of $\sigma_x$ eigenstates in the two-state
system. The effect observed here is also consistent with a full
thermodynamic analysis of the spin-boson model \cite{weiss12}.

\section{Discussion}

The stochastic Liouville--von Neumann equation in its original form
(\ref{eq:SLN}), which is a universally valid, non-perturbative
representation of open-system dynamics becomes computationally
expensive in the limit of long times. Empirically, one finds an
exponential growth $N\approx \exp(\Gamma t)$ of the sample number,
where $\Gamma$ is a rate of the same order as the physical relaxation
and decoherence rates. This problem is solved using the FMSP method.

Assuming a finite memory time $\tmem$ of the dissipative mechanism,
the finite-memory stochastic propagation given by eqs. (\ref{eq:barrho})
and (\ref{eq:breverho}) provides an alternative formulation. Here the
most important dimensionless quantity determining the required number
of samples is no longer $\approx \Gamma t$, but $\Gamma\tmem$. The
algorithmic complexity is now no longer exponential, but linear in
time $t$.

The revised method's gain in computational efficiency is typically
several orders of magnitude; it is an exponential factor in the limit
of long simulation times.
It performs exceedingly well in the regime of weak to moderate
coupling, but also covers non-perturbative settings with manageable
requirements for computational resources.

No strict separation of timescales is assumed for $\tmem$, hence
non-Markovian effects are within the scope of the method.

The FMSP method is applicable whenever an open-system problem is posed
in terms of a system-reservoir coupling and spectral reservoir
characteristics and a non-perturbative approach is preferred over the
standard Born-Markov-rotating-wave approximation.

The capacity to include arbitrary driving, including sudden unitary
transformations, in the presence of system-reservoir correlations also
makes the method a candidate for simulations in the field of
multidimensional spectroscopy, where the method of hierarchic
auxiliary density matrices is established \cite{ishiz06}. A
combination of hierarchic and stochastic methods \cite{zhou05} may be
of benefit here.

\revision{%
In its present form, FMSP offers little benefit in the case of a
sluggish bath. Related work on a different type of projector, which
sets the off-diagonal elements of $\bar\rho$ to zero in addition to
the stochastic average, is currently under way. This appears to be
suitable when strong reservoir fluctuations lead to short
decoherence times. The sub-ohmic case of the spin-boson model
\cite{ander07,winte09,kast13} might be a test case for this variant.}
Evaluating eq. (\ref{eq:zwanzig}) with this type of projector \revision{also}
establishes links to the diagrammatic NICA expansions of path
integrals \cite{egger94b,thorw01a}.

\revision{%
The combined features of long-time propagation and arbitrary driving
are welcome for simulation tasks in the emerging field of
quantum engineering, in particular,} the simulation of quantum heat
engines. For its simplified SLED form \cite{stock99}, the
compatibility of the stochastic approach with optimal control has
already been proven in a thermodynamic context \cite{schmi11}.

The stochastic approach can also describe fluctuations in the
heat transfer to a thermal reservoir \cite{schmi15} and can therefore
be considered a valuable tool when studying the link between dynamics
and novel concepts \cite{brand15} in the thermodynamics in the quantum
regime.

\acknowledgments Stimulating conversations with Joachim Ankerhold and
Michael Wiedmann are gratefully acknowledged. This work was supported
by Deutsche Forschungsgemeinschaft through grant AN336/6-1.

%


\providecommand{\noopsort}[1]{}

\end{document}